\documentstyle[12pt,aps,preprint]{revtex}
\newcommand{\spone}{0.9}

\newcommand{\singlespace}{\edef\baselinestretch{\spone}\Large\normalsize}

\newcommand{\leqsim}{\,\raisebox{-1mm}{$\stackrel{{\textstyle < }}
                                           {{\textstyle \sim}}$}\,}

\begin{document}
\title{ Additive renormalization of the specific heat of O($n$) symmetric \\
systems in three-loop order }
\author{ Stuart S.~C.~Burnett}
\address{Department of Physics, University of Manitoba,
         Winnipeg, Manitoba, Canada R3T 2N2.}
\date{\today}
\maketitle
\begin{abstract}
We present three-loop formulas for the additive renormalization constant 
$A(u,\epsilon)$ and associated renormalization group function $B(u)$ for 
the specific heat of the O($n$) symmetric $\phi^{4}$ model. Using this 
result, we obtain also the amplitude function above $T_{C}$ within the 
minimally renormalized theory at fixed $d=3$.
At the fixed point, the three-loop correction to $B(u)$ turns out to be 
small (about 3\% for $n=2$).  We note that a correction of this 
size may become important at the level of accuracy expected in future 
experiments.  \\[6pt]
\end{abstract}
%

Field-theoretic renormalization group (RG) calculations based on (Borel) 
resummations of several orders of perturbation theory have yielded accurate 
predictions for the critical exponents \cite{zinn-justin89} and for many of
the universal amplitude ratios \cite{privman91} of O($n$) symmetric systems.
For $n>1$, however, similar predictions for amplitude ratios involving
quantities defined below $T_{C}$ are not available since the 
relevant perturbation series have not yet been extended to sufficiently high 
order for resummations to be effective. It has been pointed out
recently \cite{burnett97}, that
such higher order calculations for the amplitude functions of the
specific heat and superfluid density would be needed for a
fully quantitative test of unversality along the $\lambda$-line of $^{4}$He.
Another quantity which enters the formulas for the amplitude 
ratios and which has, so far, been computed only in low order is the 
renormalization group function $B(u)$ associated with the additive part 
of the renormalization of the specific heat. Like the amplitude functions,
this function has additional relevance for the analysis of experimental 
data in the nonasymptotic region \cite{dohm84,dohm85,dohm87,Cnote}.

On the basis of specific heat measurements taken in earth
orbit, Lipa {\em et al} \cite{lipa96} have shown that the rounding of data
near the $\lambda$-transition of $^{4}$He due to gravity-induced pressure 
gradients \cite{ahlers91}, can be avoided for reduced temperatures as small 
as $t\simeq 10^{-9}$. Previously, these effects restricted 
the range of useful data to temperatures $10^{-6}\leqsim t\leqsim 10^{-2}$ 
implying the need for theoretical constraints in the analysis
\cite{singsaas84}. An unconstrained fit to the data in
Ref.~\cite{lipa96} yielded the critical exponent value $\alpha=-0.01285(38)$
and an estimate $A^{+}/A^{-}=1.054(1)$ for the ratio of leading amplitudes. 
The uncertainty in this value of
$\alpha$ is smaller than that of the ``best'' RG prediction \cite{albert82}
$\alpha=-0.016(6)$ by about an order of magnitude---a clear call
for greater theoretical accuracy.
Further experiments in reduced gravity have been planned~\cite{NASA_95}.

The purpose of this note is to examine the three-loop approximation to $B(u)$
from the standpoint of the minimally renormalized $\phi^{4}$ theory in fixed
dimension \cite{dohm85,schloms89,krause90,schloms90,halfkann92}. 
This function is expected to deviate from its leading order approximation 
$B(u)\simeq n/2$ by a small amount of O($\eta$)
\cite{dohm85,krause90,nicoll85}, where $\eta$ is 
the exponent describing the decay of spatial correlations at $T=T_{C}$. 
While the role played by $B(u)$ is expected to be a minor one 
relative to the exponents and amplitude functions
\cite{dohm85,krause90,halfkann92}, it is nevertheless of interest to know 
whether a correction of this size would become significant at the higher
level of accuracy expected in reduced gravity experiments.

\begin{figure}
\begin{picture}(400,300)(0,130)
\includegraphics{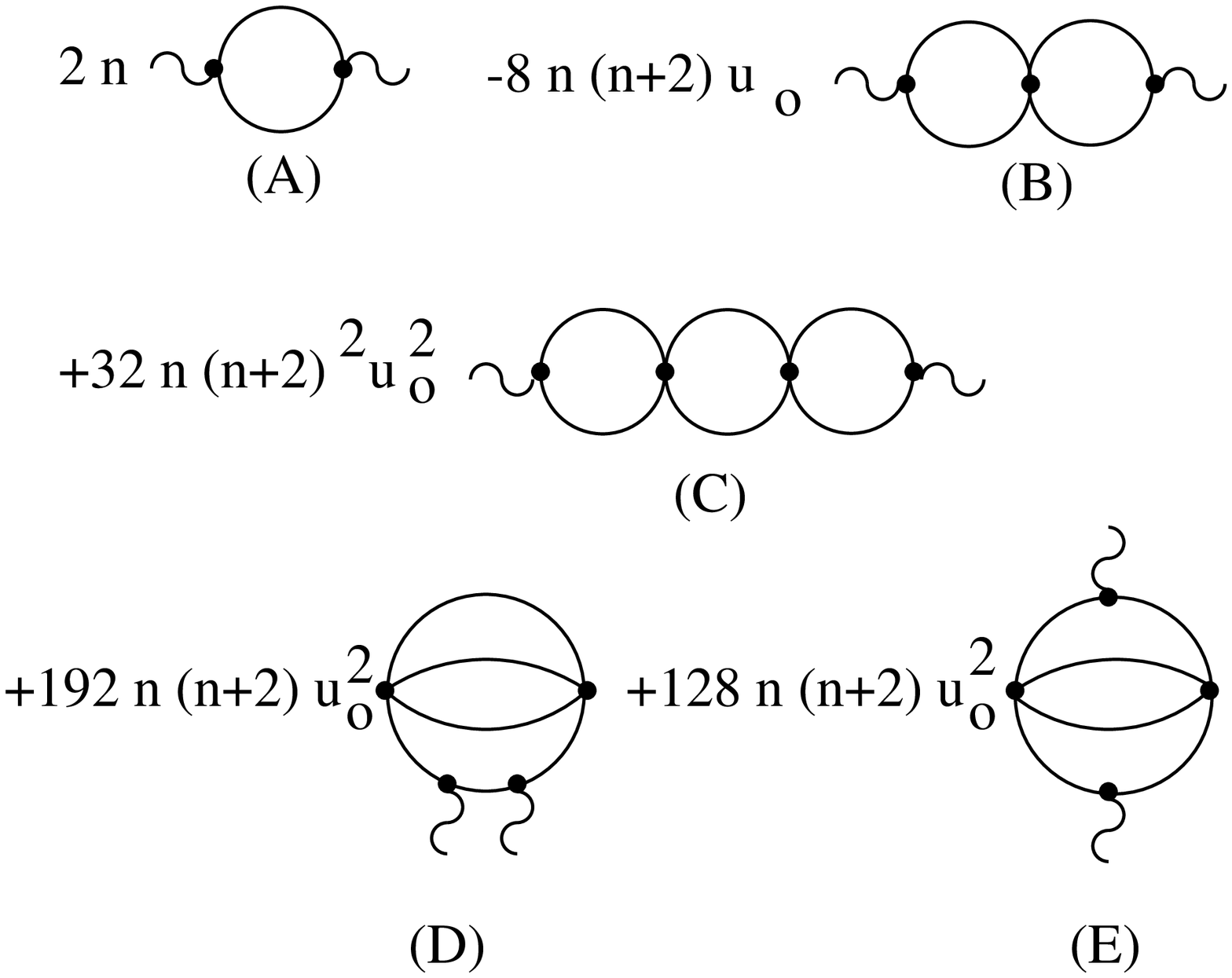}
\end{picture}
\caption{ Diagrams of the massless theory contributing to the 
specific heat. Their expansions near $d=4$ are given in
Eqs.~(\ref{eq:ia})--(\ref{eq:ie}). } \label{fig:diagrams}
\end{figure}

The renormalization of the specific heat within the minimal
subtraction scheme at fixed dimension has been described in detail
in Ref.~\cite{schloms90}. For definitions and notation, see also
Refs.~\cite{dohm85,schloms89,krause90,schloms90,halfkann92}.
In three-loop order, the additive renormalization $A(u,\epsilon)$ and 
RG function $B(u)$ are given by
\begin{eqnarray}
A(u,\epsilon) &=& -2n\frac{1}{\epsilon}-8n(n+2)\frac{u}{\epsilon^{2}}
+ a\frac{u^{2}}{\epsilon^{3}} +\mbox{O}(u^{3}) ,  \label{eq:A_3lp} \\
a &=&
-32n(n+2) \left[(n+4)-\frac{5}{3}\epsilon
+\frac{1}{8}\epsilon^{2}\right] , \nonumber \\
B(u) &=& \frac{n}{2}\,\left[ 1 + 6(n+2)u^{2} + \mbox{O}(u^{3}) \right],
\label{eq:B_3lp}
\end{eqnarray}
which we have obtained within the ``massless'' theory (that is, for $k\neq 0$ 
and $T=T_{C}$) since the pole terms are more readily evaluated there. 
The relevant vacuum diagrams with two $\phi^{2}$ insertions are shown in 
Fig.~\ref{fig:diagrams}; their contributions near $d=4$ are
\begin{eqnarray}
I_{A}&=& 2nA_{d}k^{-\epsilon}\frac{1}{\epsilon}
\left[1+ \frac{\epsilon}{2}+ [2-\zeta(2)]\frac{\epsilon^{2}}{4}
+\mbox{O}(\epsilon^{3}) \right], \label{eq:ia} \\
I_{B}&=& -8n(n+2)u_{0}A_{d}^{2}k^{-2\epsilon}
\frac{1}{\epsilon^{2}}\left[1+\epsilon+[5-2\zeta(2)]\frac{\epsilon^{2}}{4}
+\mbox{O}(\epsilon^{3}) \right], \label{eq:ib}  \\
I_{C}&=&32n(n+2)^{2}u_{0}^{2}A_{d}^{3}k^{-3\epsilon}
\frac{1}{\epsilon^{3}}\left[1+\frac{3}{2}+[9-3\zeta(2)]\frac{\epsilon^{2}}{4}
+\mbox{O}(\epsilon^{3}) \right], \label{eq:ic}  \\
I_{D}&=&-\frac{16}{3}n(n+2)u_{0}^{2}A_{d}^{3}k^{-3\epsilon}
\frac{1}{\epsilon^{2}}\left[1+\frac{15}{4}\epsilon
+\mbox{O}(\epsilon^{2}) \right], \label{eq:id}  \\
I_{E}&=&64n(n+2)u_{0}^{2}A_{d}^{3}k^{-3\epsilon}
\frac{1}{\epsilon^{3}}\left[1+2\epsilon+[13-3\zeta(2)]\frac{\epsilon^{2}}{4} 
+\mbox{O}(\epsilon^{3}) \right]. \label{eq:ie} 
\end{eqnarray}
To obtain $I_{E}$, we have used Eq.~(2.20) of Ref.~\cite{chetyrkin80}.
The geometric factor $A_{d}=2^{2-d}\pi^{-d/2}\Gamma(3-d/2)/(d-2)$ 
is left unexpanded \cite{dohm85}. 
In three dimensions, our formula for the amplitude function above $T_{C}$
reads
\begin{eqnarray}
F_{+}(1,u,3) &=& -n -2n(n+2)u + bu^{2} +\mbox{O}(u^{3}) , \label{eq:F_3lp} \\
b &=& \mbox{}-4n(n+2)\left[n+4\ln\frac{4}{3} -\frac{7}{27}\right] .
\nonumber
\end{eqnarray}

For $n=2$, the O($u^{2}$) term in Eq.~(\ref{eq:B_3lp}) is roughly 3\% of
the leading term and, since $\eta\simeq 0.04$ \cite{schloms89}, is consistent
with the O($\eta$) 
estimate of Ref.~\cite{nicoll85} for the net contribution of all higher order 
terms. It has been suggested \cite{dohm85,krause90,nicoll85} that the terms 
beyond leading order should contribute less than 1\% to the function 
$B(u)$ and yet, although this contribution is expected to be small, it is not 
at all clear that it should be so small.
One should bear in mind that low order perturbative expressions,
such as Eq.~(\ref{eq:B_3lp}), cannot by themselves be regarded as reliable 
in a purely quantitative sense and that it is usually difficult to
anticipate which (low) order of perturbation theory will provide the
``best'' approximation in any given situation. Indeed, this is the
motivation behind the resummations of higher order series that have so far
yielded accurate predictions for the exponents and amplitude ratios.

With the above {\em caveat}, therefore, let us consider the O($u^{2}$) term 
in Eq.~(\ref{eq:B_3lp}) to be O($\eta$) and examine its effect on the 
amplitude ratios \cite{schloms90}
\begin{eqnarray}
\frac{A^{+}}{A^{-}} &=& \left[\frac{2Q_{+}^{*}}{Q_{-}^{*}}\right]^{\alpha}
\frac{\alpha F_{+}^{*} + 4\nu B^{*}}{\alpha F_{-}^{*} + 4\nu B^{*}} ,
\label{eq:A+_A-}   \\
R_{\xi}^{T} &=& \left[\frac{Q_{-}^{*}}{4}\right]^{2/3}
\!\!\!\!(\alpha F_{-}^{*}+4\nu B^{*})^{1/3} \,
\frac{(4\pi)^{2/3}}{G^{*}} , \label{eq:RxiT}
\end{eqnarray}
where $Q_{\pm}$, $F_{-}$ and $G$ are the amplitude functions for the 
correlation lengths, the specific heat below $T_{C}$ and the superfluid 
density, respectively; the asterisk denotes fixed point values. 
We make use of the Borel summation results given in 
Refs.~\cite{schloms89,krause90,halfkann92} and of
the relations $Q_{+}^{*}=2\nu P_{+}^{*}$ and $Q_{-}^{*}=3-2Q_{+}^{*}$
where $P_{+}$ is the amplitude function for the quantity
$(\partial r_{0}/\partial\xi^{-2})_{u_{0}}$ \cite{schloms89}.
In the absence of Borel results for $n>1$ below $T_{C}$, we use the most 
reliable low order approximations for $F_{-}$ and $G$, which turn out to be
given already in one-loop order \cite{burnett97}.
We also set $u^{*}=0.0405$, $\alpha=0.11$ ($n=1$) and $u^{*}=0.0362$, 
$\alpha=-0.013$ ($n=2$) and fix $\nu$ according to $\nu=(2-\alpha)/3$. 

The values of $A^{+}/A^{-}$ and $R_{\xi}^{T}$ given in 
Table~\ref{tab:amp_ratios} illustrate the size and direction of the 
effect of including the O($u^{2}$) term in Eq.~(\ref{eq:B_3lp}) when all 
other quantities in Eqs.~(\ref{eq:A+_A-}) and~(\ref{eq:RxiT}) are kept fixed.
Also shown in the table is the value for $A^{+}/A^{-}$ ($n=1$) 
obtained by Bagnuls {\em et al} \cite{bagnuls87}, who use a different 
renormalization scheme, and the experimental values for $^{4}$He for 
$A^{+}/A^{-}$ and $R_{\xi}^{T}$ obtained by Lipa {\em et al} \cite{lipa96} 
and by Singsaas and Ahlers \cite{singsaas84}, respectively. 
In each case, the effect of the O($u^{2}$) term in Eq.~(\ref{eq:B_3lp}) is
comparable to the uncertainties given by the authors of
Refs.~\cite{lipa96,singsaas84,bagnuls87}.
Since the possibility of the exact value of $B^{*}$ differing from the 
leading term by $\sim 3$\% cannot be ruled out and since the
experimental uncertainties are expected to be substantially smaller in
the future, it seems that a higher order calculation, of the
kind indicated by the analysis of Ref.~\cite{burnett97} for amplitude 
functions below $T_{C}$, may be needed for $B(u)$. This conclusion, of
course, presumes the future availability of improved estimates for the
critical exponent $\alpha$ and for the amplitude functions below $T_{C}$.

It may be argued that an additive renormalization is unnecesary if the
specific heat is represented in terms of its temperature derivative
$\partial C^{\pm}/\partial t$. In this connection, we recall the relations
\cite{schloms90}
\begin{equation}
8A_{d}^{-1}P_{\pm}f_{\pm}^{(3,0)}=(\epsilon-2\zeta_{r})F_{\pm} + 4B
-\beta_{u}\frac{\partial F_{\pm}}{\partial u} , \label{eq:Pf30_Fpm}
\end{equation}
where $f_{\pm}^{(3,0)}$ are the amplitude functions for 
$\partial C^{\pm}/\partial t$. These formulas, for example, enable the 
amplitude ratios to be expressed in terms of $f_{\pm}^{(3,0)}$. In that
case, resummation results for $B(u)$ would be useful for an internal 
check of the theory. However, in view of the unusually large uncertainty 
associated with the present high order result for $f_{+}^{(3,0)}$
[12\% compared to $\leqsim 1$\% typically for other amplitude functions
\cite{krause90,halfkann92}], the representation based
on additive renormalization may well turn out to be the more reliable
in quantitative applications. A higher order calculation of $B(u)$
would be needed to answer this question.

Finally, we note that in Ref.~\cite{krause90}, the higher order coefficients
in the perturbation series for $F_{+}$ were approximated by use of
Eq.~(\ref{eq:Pf30_Fpm}) with $B(u)\simeq n/2$.
This procedure neglects the {\em leading} poles of $A(u,\epsilon)$ 
beyond two-loop order [for example, the term $\sim\frac{1}{8}\epsilon^{2}$ in 
the square brackets of Eq.~(\ref{eq:A_3lp})\,].
Using Eqs.~(\ref{eq:B_3lp}) and~(\ref{eq:Pf30_Fpm}), we find that the 
resummation results for $F_{+}^{*}$ are shifted by about 2\%.
However, since $F_{+}$ enters the formulas for the amplitude ratios and 
for the analysis of experimental data only in the combination $\alpha F_{+}$,
the effect here is entirely negligible. 

In summary, we have computed, within the framework of the minimally 
renormalized $\phi^{4}$ theory at fixed dimension $d=3$, the three-loop 
correction to the additive renormalization of the specific heat, 
Eq.~(\ref{eq:A_3lp}), for systems with O($n$) symmetry. We have used this 
to determine the corresponding RG function $B(u)$ in Eq.~(\ref{eq:B_3lp}), 
and amplitude function $F_{+}$ above $T_{C}$ in Eq.~(\ref{eq:F_3lp}). 
While the neglect (within the present scheme) of the 
{\em leading} additive poles in the specific heat beyond two-loop order 
is justified in analyses based on low order perturbation theory, 
these poles may lead to a small systematic effect 
at the level of accuracy expected in future experiments 
\cite{lipa96,NASA_95}. \\[12pt]
This work was supported in part by the National Science and 
Engineering Research Council of Canada.
\singlespace
\twocolumn
\begin{table}[h]
\caption[]{Values for the amplitude ratios obtained 
using $B(u)\simeq n/2$ [1\,\&\,2 loop] and 
Eq.~(\ref{eq:B_3lp}) [3~loop]. All other quantities in 
Eqs.~(\ref{eq:A+_A-}) and~(\ref{eq:RxiT}) are held fixed as described in the
text. The other theoretical and experimental values are included to
illustrate the current level of uncertainty for these quantities.}
\label{tab:amp_ratios}
\begin{tabular}{cccc}
       &   $A^{+}/A^{-}$  &      &     $(A^{-})^{1/3}/k_{0}$   \\ \tableline
       &   $0.527$        & [1\,\&\,2 loop]  &                          \\
 $n=1$ &   $0.536$        & [3 loop]         &                          \\
       &   $0.541(14)^{a}$ &      &                          \\
       &                   &      &                          \\
       &   $1.056$         & [1\,\&\,2 loop]  &     $0.831$                \\
 $n=2$ &   $1.054$         & [3 loop]         &     $0.840$                \\
       &   $1.054(1)^{b}$  &                  &     $0.85$---$0.86^{c}$  
\end{tabular}
\begin{minipage}{2in}
$^{a}$Bagnuls {\em et al}~\cite{bagnuls87} \\
$^{b}$Lipa {\em et al}~\cite{lipa96} \\
$^{c}$Singsaas and Ahlers~\cite{singsaas84}
\end{minipage}
\end{table}
\end{document}